\documentclass[12pt]{article}
\usepackage{a4}
\usepackage{amssymb}
\usepackage{cite}
\newcommand{\be}{\begin{equation}}
\newcommand{\ee}{\end{equation}}
\newcommand{\bea}{\begin{eqnarray}}
\newcommand{\eea}{\end{eqnarray}}

\begin{document}
\thispagestyle{empty}
\def\thefootnote{\fnsymbol{footnote}}
\begin{flushright}
{\sf ZMP-HH/09-17}\\
{\sf Hamburger$\;$Beitr\"age$\;$zur$\;$Mathematik$\;$Nr.$\;$ 345}
     \vskip 2em
\end{flushright}
\vskip 2.0em
\begin{center}\Large
Canonical quantization of the WZW model with defects and Chern-Simons theory
\end{center}\vskip 1.5em
\begin{center}
Gor Sarkissian
\footnote{\scriptsize
~Email address: \\
$~$\hspace*{2.4em}sarkissian@math.uni-hamburg.de.
}
\end{center}
\begin{center}
Organisationseinheit Mathematik, \ Universit\"at Hamburg\\
Bereich Algebra und Zahlentheorie\\
Bundesstra\ss e 55, \ D\,--\,20\,146\, Hamburg
\end{center}
\vskip 1.5em
\begin{center} July 2009 \end{center}
\vskip 2em
\begin{abstract} \noindent
We perform canonical quantization of the WZW model with defects and permutation branes.
We establish symplectomorphism between phase space of WZW model with $N$ defects
on cylinder and phase space of Chern-Simons theory on annulus  
times $R$ with $N$ Wilson lines, and between phase space of WZW model with $N$ defects
on strip and Chern-Simons theory on disc times $R$ with $N+2$ Wilson lines.
We obtained also symplectomorphism  between phase space of the $N$-fold product of the WZW model 
with boundary conditions specified by permutation branes, 
and phase space of Chern-Simons
theory on sphere with $N$ holes and two Wilson lines.
\end{abstract}

\setcounter{footnote}{0}
\def\thefootnote{\arabic{footnote}}
\newpage

\section{Introduction}
In this paper we study WZW model with defect on a world-sheet 
with and without boundary. 
The main result of this paper is proof of symplectomorphism between
the phase space of the WZW model with defects on strip or cylinder
and moduli space of flat connections on disc or annulus respectively
with sources.

To explain our results we start by reviewing Chern-Simons (CS) gauge theory
with compact gauge group $G$ on three-dimensional manifold of the form $S^2_{n,m}\times R$,
where $R$ is time direction, and $S^2_{n,m}$ is two-dimensional sphere $S^2$ with $m$ holes,
and $n$ time-like Wilson lines \cite{Elitzur:1989nr,Witten:1988hf}.
Later we will say often briefly CS theory on  $S^2_{n,m}$ suppressing ``times R''.

It was conjectured in \cite{Elitzur:1989nr} that Hilbert space of quantized
Chern-Simons theory on $S^2_{n,m}\times R$, were $n$ time-like Wilson lines
assigned with representations $\lambda_1,\ldots \lambda_n$
must be of the form
\be\label{hislop}
{\cal H}=\sum_{\tau_1,\ldots \tau_m}V_{\lambda_1,\ldots \lambda_n,\tau_1,\ldots \tau_m}\otimes 
H_{\tau_1}\otimes\ldots H_{\tau_m}
\ee
where $H_{\tau_i}$ are the representation spaces of $\hat{LG}$
corresponding to the highest weights $\tau_i$, and
 $V_{\kappa_1,\ldots \kappa_l}$ is the Hilbert space corresponding to quantizing
of Chern-Simons theory on sphere $S^2$ with $l$ Wilson lines assigned with representations 
${\kappa_1,\ldots \kappa_l}$. The latter is space of conformal blocks of the WZW model with group $G$
with dimension
\be\label{dimbl}
{\rm dim}V_{\kappa_1,\ldots \kappa_l}=\sum_{\mu_1,\ldots \mu_{l-3}}
N_{\kappa_1\kappa_2}^{\mu_1}N_{\mu_1\kappa_3}^{\mu_2}\cdots N_{\mu_{l-3}\kappa_{l-1}}^{\kappa_l}
\ee
where $N_{\nu\mu}^{\lambda}$ are fusion coefficients of the WZW model with group $G$.

Let us now compare formulas (\ref{hislop}) and (\ref{dimbl})
with different partition functions of WZW model.
The diagonal torus partition function of RCFT
is 
\be\label{zdiag}
Z=\sum_{i}\chi_i(q)\bar{\chi}_{i^*}(\bar{q}),\;\;\; 
q=\exp(2i\pi\tau)
\ee

Comparing it to (\ref{hislop}) and (\ref{dimbl}) we see that it corresponds to Hilbert space of CS theory on annulus.
This observation was made in \cite{Elitzur:1989nr}.
Using the Lagrangian of the WZW model \cite{Witten:1983ar}, it was proved in \cite{Chu:1991pn,Falceto:1992bf,Gawedzki:1990jc} that classical symplectic phase spaces 
of the WZW model on circle indeed coincides with  symplectic phase space of the CS theory
on annulus. 

The annulus partition function between Cardy  states corresponding to primaries $a$ and $b$
is

\be\label{deffuopa}
Z_{ab}={\rm Tr}_{H_{ab}}\left(\tilde{q}^{L_0-{c\over 24}}\right)=
\sum_{i}N^a_{ib}\chi_{i}(q)
\ee
Comparing it to (\ref{hislop}) and (\ref{dimbl}) we see that it corresponds to 
Hilbert space of CS theory on a disc with two Wilson lines.
Using the Lagrangian formulation of the WZW model on a world-sheet with boundary suggested in
\cite{Alekseev:1998mc,Klimcik:1996hp,Gawedzki:1999bq},
it was proved in \cite{Gawedzki:2001rm} that classical symplectic 
phase space of the WZW model on a strip coincides with symplectic phase space of CS theory 
on a disc with two Wilson lines.  
 
Now we will show that inclusion of defects 
\cite{Petkova:2000ip,Petkova:2001ag,Graham:2003nc,Fuchs:2002cm} 
and permutation branes \cite{FigueroaO'Farrill:2000ei,Recknagel:2002qq,Gaberdiel:2002jr,Fuchs:2003yk,Sarkissian:2003yw}
allows to generalize
these results to include also the following three situations:
\begin{enumerate}
\item CS theory on annulus with arbitrary number of Wilson lines,
\item CS theory on disc with arbitrary number of Wilson lines,
\item CS theory on sphere with two Wilson lines and arbitrary number of holes.
\end{enumerate}

The torus partition function with insertion of a defect  $X_a$ corresponding
to primary $a$ is given by formula:
\be\label{deffu}
Z_{a}={\rm Tr}\left(X_a\tilde{q}^{L_0-{c\over 24}}\tilde{\bar{q}}^{\bar{L}_0-{c\over 24}}\right)=
\sum_{i\bar{i}}N^a_{i\bar{i}}\chi_{i}(q)\chi_{\bar{i}}(\bar{q})
\ee

The comparison  of (\ref{deffu}) with formulas (\ref{hislop}) and (\ref{dimbl})
reveals that Hilbert space of WZW model with one defect coincides with
Hilbert space of Chern-Simons theory on annulus with one Wilson line.
Using defect fusion rule
\be\label{dds}
X_aX_b=\sum_c N^c_{ab}X_c
\ee
the formula (\ref{deffu}) can be generalized to the insertion of $N$ defects:
the torus partition function with insertion of $N$ defects corresponding
to primaries $a_i$ is 
\be
Z_{a_1\ldots a_N}=\sum_{i,\bar{i}}{\rm dim} V_{a_1\ldots a_N,i,\bar{i}}\chi_{i}(q)\chi_{\bar{i}}(\bar{q})
\ee
implying
that Hilbert space of WZW model with $N$ defects coincides with Hilbert
space of Chern-Simons theory on annulus with $N$ Wilson lines.

Using the fact that defects can be fused with boundary states producing new
boundary states
\be\label{dbs}
X_a|b\rangle=\sum_d N^d_{ab}|d\rangle\ ,
\ee
 one can compute
the annulus partition function between Cardy states corresponding to primaries 
$a$ and $b$
with insertion of a defect corresponding to primary $c$:
\be\label{deffuop}
Z_{ab,c}={\rm Tr}_{H_{ab}}\left(X_c\tilde{q}^{L_0-{c\over 24}}\right)=
\sum_{d,i}N^d_{bc}N^a_{id}\chi_{i}(q)
\ee

Comparison with (\ref{hislop}) and (\ref{dimbl}) shows that Hilbert space
of WZW model on annulus with defect coincides with Hilbert
space of Chern-Simon theory on disc with three Wilson lines.
This result can be generalized to the insertion of any number $N$ of defects as well:
the annulus partition function between Cardy states corresponding to primaries $a$ and $b$
with insertion of $N$ defects corresponding to primaries $d_i$ is
\be
Z_{ab,d_1\ldots d_N}=\sum_i{\rm dim}V_{ab,d_1\ldots d_N,i}\chi_{i}(q)
\ee

It corresponds to Chern-Simons theory on disc with $N+2$ Wilson lines.

The annulus partition function between two permutation branes 
on two-fold product of the WZW models,
corresponding
to single copy primaries $a_1$ and $a_2$ is
\be\label{perper}
Z_{a_1,a_2}= \sum_{r,k,l} N^{a_2}_{a_1r}N^r_{kl}\chi_{k}(q)\chi_{l}(q)
\ee
Partition function (\ref{perper}) corresponds to CS theory on annulus with two Wilson lines.
Again (\ref{perper}) can be generalized for permutations branes on $N$-fold product:
the annulus partition function between two permutation branes corresponding
to single copy primaries $a_1$ and $a_2$ on $N$-fold product is
\be\label{perN}
Z_{a_1,a_2}=\sum_{i_1,\ldots i_N}{\rm dim}V_{a_1,a_2,i_1,\ldots i_N}\chi_{i_1}(q)\ldots\chi_{i_N}(q)
\ee
Partition function (\ref{perN}) corresponds to CS theory on sphere with $N$ holes and two Wilson lines.

Given these results it is natural to assume that classical symplectic phase
spaces of the WZW model with defects, suggested in \cite{Fuchs:2007fw} 
and with permutation branes suggested in \cite{Sarkissian:2003yw}
should coincide 
with simplectic phase  space of the CS theory in the mentioned situations as well. In this paper we prove that it is the case.

The paper is organized in the following way. In the section 2 we review
symplectic form on moduli space of flat connections on $S^2_{m,n}$.
In section three we review symplectic phase space of WZW model on a cylinder.
In section four we review symplectic phase space of WZW model on a strip.
After all these preparations in section five we present symplectic phase
space of WZW models on cylinder and strip with $N$ defects and show
that they have the same structure and symplectic form as 
symplectic moduli space of flat connections on annulus with $N$ sources and 
disc with $N+2$ sources respectively.
In the last section we show that symplectic phase space of the $N$-fold product of WZW models
on strip with boundary conditions specified by permutation branes  coincides 
with symplectic moduli space of flat
connections on sphere with $N$ holes and with two sources.

\section{Moduli space of flat connections on $S^2_m$}

Here we present details on symplectic form on moduli space 
of flat connections on sphere $S^2_{n,m}$ 
with $n$ Wilson line and $m$ holes.

Let us at the beginning recall some essential points 
on the CS theory with Wilson lines.

It was shown in \cite{Elitzur:1989nr,Witten:1988hf} that phase space
of CS theory on manifold of form $M\times R$, where $M$ is two-dimensional
Riemann surface, $R$ is time direction,
with $n$ time-like Wilsonian lines assigned with representations $\lambda_i$,
is moduli space ${\cal P}_{S^2_{n,m}}$ of connections $A$ on $M$ satisfying the equation:
\be\label{mommap}
{k\over 2\pi}F(z)+ i\sum_{i=1}^nT_i\delta(z-z_i)=0
\ee
where $F=dA+A^2$ and  $z_i$ are points where Wilson lines hit $M$.
$T_i$ are conjugacy classes in the Lie algebra ${\rm g}$
\be
T_i=v_i\lambda_i v_i^{-1},\;\;\;\; v_i\in G
\ee
where $\lambda_i$ take values in the Cartan subalgebra. 
Recall also the following remark \cite{Elitzur:1989nr,Gawedzki:2001rm}.
The symplectic form on moduli space of flat connections on sphere with $n$ sources and $m$ holes
can be decomposed as sum of symplectic forms on moduli space of flat connections
on sphere $S^2_{n+m,0}$ with $n+m$ sources and $m$ copies of the symplectic form on moduli space of flat connections  on the two-dimensional
disc with one source $D_1$:
\be\label{decsn}
\Omega_{S^2_{n,m}}=\Omega_{S^2_{n+m,0}}+\sum_{i=1}^m\Omega_{D_{1_i}}
\ee
Decomposition (\ref{decsn}) implies that to write symplectic form 
on ${\cal P}_{S^2_{n,m}}$ it is enough to know symplectic form 
on ${\cal P}_{S^2_{n,0}}$ and ${\cal P}_{D_{1}}$.

The symplectic form on the moduli space of flat connections on 2-dimensional manifold $M$ with $n$ sources is given by formula
\be\label{omnh}
\Omega={k\over 4\pi}{\rm tr}\int_{M} (\delta A)^2+i \sum_{i=1}^n {\rm tr}(\lambda_i(v_i^{-1}\delta v_i)^2)
\ee
where $A$ satisfies (\ref{mommap}). The $\delta$ denotes here exterior derivative on
moduli space.

For the case of disc with one source (\ref{omnh}) takes form:
\be\label{omd1}
\Omega_{D_1}={k\over 4\pi}\int_{D} {\rm tr}(\delta A)^2+ i[{\rm tr}(\lambda(v^{-1}\delta v)^2)]
\ee
and 
solution of (\ref{mommap}) is:
\be\label{a1}
A=-{i\over k}\eta\lambda\eta^{-1}d\phi-d\eta\eta^{-1}
\ee
where $\phi$ is angular coordinate on the disc,  $\eta\in G$ is single-valued on the disc and $\eta(z_1)=v$.

To calculate (\ref{omd1}) it was proved in \cite{Alekseev:1993rj} the following useful lemma:
Suppose that
\be
A=\eta B\eta^{-1}-d\eta\eta^{-1}
\ee
where $B$ is a gauge field and $\eta\in G$.
Then 
\be
\omega={\rm tr}\int_{D} (\delta A)^2
\ee
can be written as
\be
\omega={\rm tr}\int_{D}\{(\delta B)^2-2\delta[F_B\eta^{-1}\delta\eta]\}+
{\rm tr}\int_{\partial D}\{\eta^{-1}\delta\eta d(\eta^{-1}\delta\eta)+
2\delta[B\eta^{-1}\delta\eta]\}
\ee
where $F_B=dB+B^2$. One can prove this lemma by straightforward calculation.

Using this lemma one can easily obtain
\be\label{omdisc}
\Omega_{D_1}=\int_{\partial D} {k\over 4\pi}{\rm tr}(\eta^{-1}\delta\eta)d(\eta^{-1}\delta\eta)
+{1\over 2\pi}{\rm tr}(i\lambda(\eta^{-1}\delta\eta)^2)d\phi
\ee
It is shown in \cite{Pressley:1988qk} that geometrical quantization of 
the coadjoint orbits of $\hat{LG}$ with this form leads to the integrable representation $H_{\lambda}$ of the affine algebra
$\hat{\rm g}$ at level $k$.
 By this reason later we denote this form $\Omega^{\rm LG}(\eta,\lambda)$.
 
To compute symplectic form on the sphere with sources
it was suggested in \cite{Alekseev:1993rj} the following strategy. 
We choose a reference point $P_0$ on sphere and draw loops $l_i$ around each source point $z_i$ starting and 
ending at the chosen reference point $P_0$.
After that we cut out a sphere along these loops.
After this operation we have $n$ discs $D_i$ centered around sources $z_i$, and each of which having
as boundary one of these loops $\partial D_i=-l_i$,  and additionally  
disc $D_0$ whose boundary formed by the sum of all of them: $\partial D_0=\sum_{i=1}^n l_i$.

Introducing  local angular coordinate $\phi_i$ on discs $D_i$ around point $z_i$ one can locally write
as before (\ref{a1})

\be\label{ai}
A_i=-{i\over k}\eta_i\lambda_i\eta^{-1}_i d\phi_i-d\eta_i\eta_i^{-1}
\ee
\be
\Omega^{\rm Disc}_i=\int_{l_i} {k\over 4\pi}{\rm tr}(\eta_i^{-1}\delta\eta_i)d(\eta_i^{-1}\delta\eta_i)
+{1\over 2\pi}{\rm tr}(i\lambda_i(\eta_i^{-1}\delta\eta_i)^2)d\phi_i
\ee
The solution (\ref{ai}) implies that holonomy $M_i$ of flat connection around
point $z_i$ takes value  in conjugacy classes ${\cal C}_i$: 
\be
M_i=\eta_ie^{2\pi i\lambda_i/k}\eta_i^{-1}
\ee

On disc $D_0$ there are no sources and one has usual flat connection
\be
A_0=-d\eta_0\eta_0^{-1},\;\;\;\;\; \eta_0\in G
\ee
The corresponding symplectic form again easily derived form the lemma above:
\be
\Omega_0={k\over 4\pi}\int_{\partial D_0}{\rm tr}\{\eta^{-1}_0\delta\eta_0 d(\eta^{-1}_0\delta\eta_0)\}
\ee
Now for symplectic form one can write
\bea\label{oms22}
&&\Omega_{S^2_{n,0}}=
\Omega_0+\sum_{i=1}^n\Omega_i=\\ \nonumber
&&{k\over 4\pi}\sum_{i=1}^n\int_{l_i}{\rm tr}\{\eta^{-1}_0\delta\eta_0 d(\eta^{-1}_0\delta\eta_0)-
\eta_i^{-1}\delta\eta_id(\eta_i^{-1}\delta\eta_i)
-{2i\over k}\lambda_i(\eta_i^{-1}\delta\eta_i)^2d\phi_i\}
\eea
The last thing which we should do is to match connection $A_0$ with connections $A_i$
along the boundaries of $D_0$ and $D_i$:
\be\label{mati}
A_0|_{l_i}=A_i|_{l_i}
\ee
Equation (\ref{mati}) easily can be solved
\be\label{matet}
\eta_0|_{l_i}=\eta_i|_{l_i}\exp({i\over k}\lambda_i\phi_i)N_i
\ee
where $N_i$ is constant.
Denote values of $\eta_0$ at end points $p_{i-1}$ and $p_i$ of $l_i$ 
by 
\be\label{ki}
K_i=\eta_0(p_i)\;\;\;\;\;   K_{i-1}=\eta_0(p_{i-1})
\ee
Equation(\ref{matet})  implies that they satisfy 
\be\label{kkm}
K_iK_{i-1}^{-1}=M_i
\ee
Remembering that $\eta_i$ is single-valued on the disc we have also  
\be\label{ei}
\eta_i(p_i)=\eta_i(p_{i-1})
\ee
Going around full boundary of $D_0$ implies
\be\label{holcon}
 M_n\cdots M_1=1
\ee

Inserting (\ref{matet}) to (\ref{oms22}) one obtains:
\be
\Omega_{S^2_{n,0}}={k\over 4\pi}\sum_{i=1}^n{\rm tr}[N_i^{-1}\delta N_i\eta_0^{-1}\delta\eta_0]|^{p_i}_{p_{i-1}}=
-{k\over 4\pi}\sum_{i=1}^n{\rm tr}[\delta\eta_i\eta_i^{-1}\delta\eta_0\eta_0^{-1}]|^{p_i}_{p_{i-1}}
\ee
Using (\ref{ki}), (\ref{ei}) and (\ref{kkm}) finally we arrive to
\be\label{omms2}
\Omega_{S^2_{n,0}}={k\over 4\pi}\sum_{i=1}^n \omega_{\lambda_i}(M_i)
+{k\over 4\pi}\sum_{i=1}^n {\rm tr}(K_{i-1}^{-1}\delta K_{i-1}K_{i}^{-1}\delta K_{i})
\ee

\be\label{omf}
\omega_{\lambda_i}(M_i)={\rm tr}(\eta_i^{-1}\delta\eta_ie^{2\pi i\lambda_i/k}\eta_i^{-1}\delta\eta_ie^{-2\pi i\lambda_i/k})
\ee 
One can solve (\ref{kkm}) for $K_i$.  Let us remark that without loss of generality one can choose $\eta_0$
in such a way that its value $K_0$ is equal to the unity element. After that the
$K_i$ will be given by products of $M_i$:
\be
K_i=M_i\cdots M_1
\ee

At this point we should note that the derivation  above is carried out in the assumption 
that all holonomies $M_i$ taking values in conjugacy classes with fixed $\lambda_i$.
Actually what happens that some of them indeed take values in the fixed conjugacy classes,
but some of them rather should be considered as taking their values in continuous 
families of conjugacy classes, which would be reduced to discrete families 
upon quantization. This can be also understood from the formula (\ref{holcon}),
requiring the product of all holonomies to be unity. It is clear from this formula, that
one can solve for one of the holonomies, say $M_n$ in the term of product of others 
$M_n=M_1^{-1}\cdots M_{n-1}^{-1}$. But product of conjugacy classes is not a conjugacy class. 
Hence, we should decompose $M_n$ as continuous family of conjugacy classes.
Given that after quantization we obtain space of conformal blocks, this consideration
implies that the discrete family which will be derived after quantization is determined by
fusion rules.
Now we are ready to describe moduli space of flat connections on sphere with sources.
Assume we have $k$ holonomies $M_i$ with fixed conjugacy classes, and $n-k$ $M_j$
with holonomies in continuous families.
In this case 
\be\label{moduli}
{\cal P}_{S^2_{n,0}}=(M_1,\ldots, M_{k}, \eta_{k+1},\lambda_{k+1},\ldots, \eta_{n},\lambda_{n})/G
\ee
with the relation (\ref{holcon}) where $M_j=\eta_je^{2\pi\lambda_j/k}\eta_j^{-1}$, $j=k+1,\ldots, n$.
$G$ acts here by simultaneous adjoint action on $M_i$, $i=1,\ldots k$, and by left action on
$\eta_j$, $j=k+1,\ldots n$. This action is induced by the local gauge transformation of the gauge connections.
The mentioned holonomies with  continuous $\lambda$ will modify also obtained symplectic forms 
for disc and sphere.
The symplectic form on moduli space of flat connections on a disc with one source
(\ref{omdisc}) will take form
\be\label{omdiscm}
\Omega^{\rm LG}(\eta,\lambda)={k\over 4\pi}\int_{\partial D} {\rm tr}(\eta^{-1}\delta\eta)d(\eta^{-1}\delta\eta)
+{2\over k}{\rm tr}(i\lambda(\eta^{-1}\delta\eta)^2)d\phi-{2\over k}{\rm tr}(i\delta \lambda\eta^{-1}\delta \eta) d\phi
\ee
and the form (\ref{omms2}) will be modified by the following term:
\be\label{modterm}
\sum_{j=k+1}^n{\rm tr}(i\delta\lambda_j\eta^{-1}_j\delta \eta_j)
\ee

Let us briefly explain how quantization of the  moduli space of flat connection on $S^2_{n,0}$
with form (\ref{omms2})
leads to the space of conformal blocks considered in introduction.
Another important result obtained in \cite{Alekseev:1993rj} is that by a change
of variables symplectic form (\ref{omms2}) can be written as sum of $\Omega^{PL}$ forms,
\be
\Omega_{S^2_{n,0}}=\sum_{i=1}^n\Omega^{PL}(M_i)
\ee
where
\be 
\Omega^{PL}(M)=\omega_{\lambda}(M)+L_+^{-1}\delta L_+L_-^{-1}\delta L_-
\ee
$L_+$ and $L_-$ here are components of the Gauss decomposition $L_+L_-=M$.
On the other side it is known that  quantization with $\Omega^{PL}$ leads
to the highest weight representations $\Upsilon_{q,\lambda}$ of the deformed enveloping algebra
${\cal U}_q({\rm g})$ \cite{Alekseev:1993qs,Falceto:1992bf,SemenovTianShansky:1985my}. Hence quantizing 
${\cal P}_{S^2_{n,0}}$ with the form $\Omega_n$ leads to the tensor
product $\otimes_i\Upsilon_{q,\lambda_i}$. Gauge transformation of gauge connections give
rises on the quantum level to the diagonal action of ${\cal U}_q({\rm g})$ on $\otimes_i\Upsilon_{q,\lambda_i}$.
Therefore, in the first approximation, we obtain the subspace of invariant tensors
of that action. More precisely, the subspace of invariants may be equipped with a semipositive scalar product and one should divide by the subspace of null-vectors. The quotient spaces are isomorphic
to the spaces of conformal blocks of the WZW theory. 

Combining this with comment after formula (\ref{omdisc})
and decomposition (\ref{decsn}) we obtain (\ref{hislop}).

We finish this section by writing explicitly  formula (\ref{omms2}) for the cases $n=3$ and $n=4$,
which we need in next sections.
For the case of $n=3$
\be\label{om3}
\Omega_{S^2_{3,0}}={k\over 4\pi}\sum_{i=1}^3 \omega_{\lambda_i}(M_i)+{k\over 4\pi}{\rm tr}(\delta M_1M_1^{-1}M_2^{-1}\delta M_2)
\ee

For the case of $n=4$ the second term in (\ref{omms2}) can be written in two equivalent forms:
\be\label{om4a}
\Omega_{S^2_{4,0}}={k\over 4\pi}\sum_{i=1}^4 \omega_{\lambda_i}(M_i)
+{k\over 4\pi}{\rm tr}(\delta M_1M_1^{-1}M_2^{-1}\delta M_2+\delta M_3M_3^{-1}M_4^{-1}\delta M_4)
\ee
or
\be\label{om4b}
\Omega_{S^2_{4,0}}={k\over 4\pi}\sum_{i=1}^4 \omega_{\lambda_i}(M_i)+
{k\over 4\pi}{\rm tr}(\delta M_1M_1^{-1}M_2^{-1}\delta M_2+\delta M_1M_1^{-1}M_2^{-1}M_3^{-1}\delta M_3 M_2
+\delta M_2 M_2^{-1}M_3^{-1}\delta M_3)
\ee

\section{Bulk WZW model}
In this section we review canonical quantization of the
WZW model on the cylinder $\Sigma=R\times S^1=(t,x\; {\rm mod}\; 2\pi)$ 
\cite{Chu:1991pn,Falceto:1992bf,Gawedzki:1990jc}. 
The world-sheet action of the bulk WZW model is 
\be
S^{\rm bulk}(g)={k \over 4\pi}\int_{\Sigma}{\rm Tr}(g^{-1}\partial_{+} g)(g^{-1}\partial_{-}g)dx^+dx^-
 +{k \over 4\pi}\int_B \omega^{WZ}(g)
 \ee
where $x^{\pm}=x\pm t$, and
\be
\omega^{WZ}(g)={1\over 3}{\rm tr}(g^{-1}\delta g)^3
\ee
The phase space of solutions ${\cal P}$ can be described by the Cauchy data 
\footnote{ Surely we can choose any time slice, but for simplicity we always below take slice $t=0$.}
at $t=0$.
\be
g(x)=g(0,x)\;\;\; {\rm and}\;\;\; \xi_0(x)=g^{-1}\partial_tg(0,x)
\ee

The corresponding symplectic form is \cite{Gawedzki:1990jc}
\be
\Omega^{\rm bulk}={k\over 4\pi}\int_0^{2\pi}\Pi(g) dx
\ee
where 
\be\label{pigden}
\Pi(g)={\rm tr}\left(-\delta\xi_0g^{-1}\delta g+
(\xi_0+g^{-1}\partial_x g)(g^{-1}\delta g)^2\right)
\ee
The $\delta$ denotes here as before exterior derivative on the phase space ${\cal P}$.
It is easy to check that the symlectic form density $\Pi(g)$ has the following exterior derivative
\be\label{dpi}
\delta \Pi(g)=\partial_x\omega^{WZ}(g)
\ee
what implies closedness of the $\Omega$
\be
\delta\Omega^{\rm bulk}=0
\ee

The classical equations of motion are
\be\label{eqmot}
\partial_{-}J_L=0\;\;\;{\rm and} \;\;\;\partial_{+}J_R=0
\ee
where 
\be
J_L=-ik \partial_{+}gg^{-1}\;\;\;{\rm and}\;\;\; J_R=ikg^{-1}\partial_{-}g
\ee
The general solution of (\ref{eqmot}) satisfying boundary conditions 
\be
g(t,x+2\pi)=g(t,x)
\ee
is
\be\label{bulkdec}
g(t,x)=g_L(x^+)g^{-1}_R(x^-)
\ee
with $g_{L,R}$ satisfying monodromy conditions
\be\label{monl}
g_{L}(x^++2\pi)=g_{L}(x^+)\gamma
\ee
\be\label{monr}
g_{R}(x^-+2\pi)=g_{R}(x^-)\gamma
\ee
with the same matrix $\gamma$.
Expressing the symlectic form density $\Pi(g)$ in the terms of $g_{L,R}$ we obtain
\be\label{denlr}
\Pi=
{\rm tr}\left[g_L^{-1}\delta g_L\partial_x(g_L^{-1}\delta g_L)-
g_R^{-1}\delta g_R\partial_x(g_R^{-1}\delta g_R)
+\partial_x (g_L^{-1}\delta g_Lg_R^{-1}\delta g_R)\right]
\ee
Using (\ref{denlr}) and (\ref{monl}), (\ref{monr}) one derives for $\Omega$
\be\label{rldeco}
\Omega^{\rm bulk}=\Omega_L-\Omega_R
\ee
where
\be
\Omega_L={k\over 4\pi}\int_0^{2\pi}{\rm tr}\left(g_L^{-1}\delta g_L\partial_x(g_L^{-1}\delta g_L)\right)dx+
{k\over 4\pi}{\rm tr}(g_L^{-1}\delta g_L(0)\delta\gamma\gamma^{-1})
\ee
and $\Omega_R$ is given by the same formula with $g_R\rightarrow g_L$.
The chiral field $g_L$ can be decomposed into the product of a closed loop in $G$,
a multivalued field in the Cartan subgroup and a constant element in $G$:
\be\label{decomp}
g_L=h(x)e^{i\tau x/k}g_0^{-1}
\ee
where $h\in LG$, $\tau\in t$ ( the Cartan algebra) and $g_0\in G$. For the monodromy
of $g_L$ we obtain
\be
\gamma=g_0e^{2i\pi\tau/k }g_0^{-1}
\ee
Parametrization (\ref{decomp}) induces the following decomposition of $\Omega_L$
\be\label{omdecom}
\Omega_L=\Omega^{LG}(h,\tau)+{k\over 4\pi}\omega_{\tau}(\gamma)+{\rm tr}[(i\delta\tau)g_0^{-1}\delta g_0]
\ee
where $\Omega^{LG}(h,\tau)$ is the form (\ref{omdiscm}):
\be
\Omega^{LG}(h,\tau)={k\over 4\pi}\int_0^{2\pi}{\rm tr}[h^{-1}\delta h\partial_x(h^{-1}\delta h)+
{2i\over k}\tau(h^{-1}\delta h)^2
-{2i\over k}(\delta \tau)h^{-1}\delta h]dx
\ee
and $\omega_{\tau}(\gamma)$ is the same form as defined in (\ref{omf}):
\be\label{omfik}
\omega_{\tau}(\gamma)={\rm tr}[g_0^{-1}\delta g_0e^{2i\pi\tau/k}g_0^{-1}\delta g_0e^{-2i\pi\tau/k}]
\ee

Comparing (\ref{rldeco}) with (\ref{decsn}) for $n=0$ and $m=2$, we see that symplectic phase of the WZW model on 
circle coincides with that of CS theory on annulus.

\section{Boundary WZW model}

Here we review canonical quantization of the WZW model on the strip 
$M=R\times [0,\pi]$ for maximally symmetric boundary conditions \cite{Gawedzki:2001rm}.
Let us remind some well-known stuff on Lagrangian formulation of the WZW model on a
world-sheet with boundary \cite{Alekseev:1998mc,Klimcik:1996hp,Gawedzki:1999bq}.
Consider at the beginning the case when $M$ has one boundary.
It is well established that maximally symmetric boundary conditions:
\be\label{curbound}
J_L=-J_R|_{\partial M}
\ee
requires fields on boundary take values in discrete set of  conjugacy classes:
\be\label{gbound}
g|_{\partial M}\in {\cal C}_{\mu}=\beta e^{2i\pi\mu/k}\beta^{-1},\;\;\;\; \beta\in G
\ee
where $\mu\equiv${\boldmath $\mu\cdot H$} is a highest weight representation integrable at level $k$,
taking value in the Cartan subalgebra.

To write down action of WZW model one should choose auxiliary disc $D$ satisfying
condition $\partial B=M+D$, and continue $g$ on that disc always taking values in conjugacy class.
With such a set-up action takes form
 \be\label{actbound}
S^{\rm boundary}=S^{\rm bulk}-{k\over 4\pi}\int_{D} \omega_{\mu}
\ee
where
$\omega_{\mu}$ is the form
defined in (\ref{omf}). This form satisfies the condition 
\be\label{ommu}
\omega^{\rm WZW}(g)|_{g\in {\cal C}_{\mu}}=d\omega_{\mu}
\ee
which guaranties that the action (\ref{actbound}) is well defined.
In the case of several boundaries the condition (\ref{gbound}) 
should be imposed on each boundary component, and the corresponding boundary two-form should be added for each component as well.

From the paragraph above follows that for the case of strip we should impose
the following boundary conditions
\be\label{gfbound}
g(t,0)\in {\cal C}_{\mu_0},\;\;\; g(t,\pi)\in {\cal C}_{\mu_{\pi}}
\ee

The solution of bulk equation of motions (\ref{eqmot}) with boundary conditions
(\ref{curbound}) is found in \cite{Gawedzki:2001rm}.
It takes again the form (\ref{bulkdec}) but with $g_{L,R}$ satisfying:
\be\label{opl}
g_{L}(y+2\pi)=g_{L}(y)\gamma,\;\;\; {\rm and} \;\;\; g_{R}(y)=g_{L}(-y)h_0^{-1}
\ee 
The equations (\ref{opl}) imply
\be
g(t,0)=g_{L}(t)g_{R}^{-1}(-t)=g_{L}(t)h_0g_{L}^{-1}(t)
\ee
and
\be
g(t,\pi)=g_{L}(\pi+t)g_{R}^{-1}(\pi-t)=g_{L}(-\pi+t)\gamma h_0g_{L}^{-1}(-\pi+t)
\ee
Therefore to be in agreement with (\ref{gfbound}) one should require
\be\label{b}
h_0\in {\cal C}_{\mu_0},\;\;\; {\rm and} \;\;\; \gamma h_0=h_{\pi} \in {\cal C}_{\mu_{\pi}}
\ee
The symplectic form on the phase space of the WZW model on the strip is: 
\be\label{omopen}
\Omega^{\rm strip}={k\over 4\pi}\left[\int_0^{\pi}\Pi(g)dx+\omega_{\mu_0}(g(0,0))
-\omega_{\mu_{\pi}}(g(0,\pi))\right]
\ee
The equations (\ref{dpi}), (\ref{ommu}) imply that the form  (\ref{omopen})
is closed. Inserting (\ref{pigden}) in (\ref{omopen}) one obtains:
\bea
{4\pi \over k}\Omega^{\rm strip}=\int_0^{\pi}{\rm tr}(g_{L}^{-1}\delta g_{L}\partial_x(g_{L}^{-1}\delta g_{L}))
dx-\int_0^{\pi}{\rm tr}(g_{R}^{-1}\delta g_{R}\partial_x(g_{R}^{-1}\delta g_{R}))dx\\ \nonumber
+{\rm tr}(g_{L}^{-1}\delta g_{L}g_{R}^{-1}\delta g_{R})(\pi)-
{\rm tr}(g_{L}^{-1}\delta g_{L}g_{R}^{-1}\delta g_{R})(0)\\ \nonumber
+\omega_{\mu_0}(g(0,0))
-\omega_{\mu_{\pi}}(g(0,\pi))
\eea
Using (\ref{opl}) we obtain:
\bea\label{r1l1g}
&&-\int_0^{\pi}{\rm tr}(g_{R}^{-1}\delta g_{R}\partial_x(g_{R}^{-1}\delta g_{R}))dx=
\int_{\pi}^{2\pi}{\rm tr}(g_{L}^{-1}\delta g_{L}\partial_x(g_{L}^{-1}\delta g_{L}))dx\\ \nonumber
&&-{\rm tr}(h_0^{-1}\delta h_0g_{L}^{-1}\delta g_{L}(0))+
{\rm tr}(h_0^{-1}\delta h_0g_{L}^{-1}\delta g_{L}(-\pi))
-{\rm tr}(\delta\gamma\gamma^{-1}g_{L}^{-1}\delta g_{L}(0))+{\rm tr}(\delta\gamma\gamma^{-1}g_{L}^{-1}\delta g_{L}(-\pi))
\eea
With the help of the following useful formula obtained in \cite{Elitzur:2001qd}:
\be\label{usfor}
\omega_f(\lambda C\lambda^{-1})=\omega_f(C)+{\rm tr}(\lambda^{-1}\delta\lambda C\lambda^{-1}\delta\lambda C^{-1})
+{\rm tr}(C^{-1}\delta C+\delta CC^{-1})\lambda^{-1}\delta\lambda
\ee
one can show
\be\label{om0}
\omega_{\mu_0}(g(0,0))-{\rm tr}(g_{L}^{-1}\delta g_{L}g_{R}^{-1}\delta g_{R})(0)
-{\rm tr}(h_0^{-1}\delta h_0g_{L}^{-1}\delta g_{L}(0))=\omega_{\mu_0}(h_0)
\ee
and
\bea\label{ompi}
-\omega_{\mu_{\pi}}(g(0,\pi))+{\rm tr}(g_{L}^{-1}\delta g_{L}g_{R}^{-1}\delta g_{R})(\pi)
+{\rm tr}(h_0^{-1}\delta h_0g_{L}^{-1}\delta g_{L}(-\pi))\\ \nonumber
+{\rm tr}(\delta\gamma\gamma^{-1}g_{L}^{-1}\delta g_{L}(-\pi))=
-\omega_{\mu_{\pi}}(\gamma h_0)+{\rm tr}(\delta h_0h_0^{-1}\gamma^{-1}\delta\gamma)
\eea
Collecting all we receive
\be
\Omega^{\rm strip}=\Omega_L+{k\over 4\pi}\left[\omega_{\mu_0}(h_0)-\omega_{\mu_{\pi}}(\gamma h_0)+{\rm tr}(\delta h_0h_0^{-1}\gamma^{-1}\delta\gamma)\right]
\ee
Finally again using for $g_L$ decomposition (\ref{decomp}) and taking into account 
(\ref{omdecom}) one obtains:
\be\label{omstrip}
\Omega^{\rm strip}=\Omega^{LG}(h,\tau)+\Omega^{\rm bndry}
\ee
where
\be\label{ombndry}
\Omega^{\rm bndry}={\rm tr}[(i\delta\tau)g_0^{-1}\delta g_0]
+{k\over 4\pi}\left[\omega_{\tau}(\gamma)+\omega_{\mu_0}(h_0)-\omega_{\mu_{\pi}}(\gamma h_0)+{\rm tr}(\delta h_0h_0^{-1}\gamma^{-1}\delta\gamma)\right]
\ee
Recalling (\ref{b}) we see that boundary phase space is
\be
{\cal P}^{\rm bndry}=(h_0, h_{\pi}, g_0,\tau)/G
\ee
subject to relation $\gamma h_0=h_{\pi}$, where $\gamma=g_0e^{2i\pi\tau/k}g_0^{-1}$.
As explained in formula (\ref{moduli}) this is moduli space of flat connections on sphere
with three Wilson lines ${\cal P}_{S^2_{3,0}}$.
The symplectic form (\ref{ombndry}) coincides with (\ref{om3}) with the term (\ref{modterm}).
Comparing (\ref{omstrip})  with (\ref{decsn}) for $n=2$ and $m=1$, we
see that symplectic phase space of the WZW model on the strip
coincides with that of CS theory on the disc with two Wilson lines.

\section{WZW model with Topological defects} 
\subsection{Closed strings}

Let us assume that one has defect line separating world-sheet on two
regions $\Sigma_1$ and $\Sigma_2$. In such a situation WZW model
defined by pair of maps $g_1$ and $g_2$.
Maximally-symmetric topological defects defined as defect lines satisfying 
conditions:
\be\label{defcur}
J_{L_1}=J_{L_2}|_{\rm defect\; line}\;\;\; {\rm and} \;\;\; J_{R_1}=J_{R_2}|_{\rm defect\; line}
\ee

It is shown in \cite{Fuchs:2007fw} that the conditions (\ref{defcur})
imply that on the defect line fields $g_1$ and $g_2$ satisfy the constraint
\be\label{defline}
g_1g^{-1}_2|_{\rm defect\; line}=F\in {\cal C}_{\mu}=\beta e^{2i\pi\mu/k}\beta^{-1},\;\;\; \beta\in G
\ee
where $\mu\equiv${\boldmath $\mu\cdot H$},  as before, is a highest weight representation 
integrable at level $k$,
taking value in the Cartan subalgebra.
To write action of the WZW model with defect one again should 
introduce auxiliary disc satisfying conditions
\be
\partial B_1=\Sigma_1+\bar{D}\;\; {\rm and}\;\; \partial B_2=\Sigma_2+D
\ee
and continue fields $g_1$ and $g_2$ on this disc always holding 
the condition (\ref{defline}).
After this preparations the action takes form \cite{Fuchs:2007fw}:
\be\label{actdef}
S=S^{\rm bulk}(g_1)+S^{\rm bulk}(g_2)+{k\over 4\pi}\int_D\varpi(g_1,g_2)
\ee
where
\be\label{varmar}
\varpi(g_1,g_2)=\omega_{\mu}(F)-{\rm Tr}(g_1^{-1}dg_1g_2^{-1}dg_2)
\ee
The form (\ref{varmar}) satisfies the equation:
\be\label{extdef}
d\varpi(g_1,g_2)=\omega^{WZ}(g_1)|_{\rm defect}-\omega^{WZ}(g_2)|_{\rm defect}
\ee
Equation (\ref{extdef}) guarantees that the action (\ref{actdef}) is well defined.

Now consider WZW model on the same cylinder as in section 1, and put defect line
at $x=a$ in parallel to the time line. 

The solution of the (\ref{eqmot}) with defect conditions 
(\ref{defcur}) is again given by (\ref{bulkdec}) on bulk for both fields, but with
$g_{L_1}$,$g_{R_1}$,$g_{L_2}$,$g_{R_2}$ satisfying the following defect conditions:
\bea
\label{lra}
g_{L_2}(y)=g_{L_1}(y)h_a^{-1}\\ \nonumber
g_{R_2}(y)=g_{R_1}(y)m_a
\eea
The equations (\ref{lra}) imply
\be
F(t,a)=g_1g_2^{-1}(t,a)=g_{L_1}(a+t)g_{R_1}^{-1}(a-t)g_{R_2}(a-t)g_{L_2}^{-1}(a+t)=
g_{L_1}(a+t)m_ah_ag_{L_1}^{-1}(a+t)
\ee
Therefore to satisfy the boundary condition (\ref{defline}) we should require 
\be
m_ah_a=d_a\in {\cal C}_{\mu_a}
\ee 
Given that we consider WZW model on cylinder we should additionally require
\be\label{closed}
g_2(t,2\pi)=g_1(t,0)
\ee
The condition (\ref{closed}) imposes the following relation on monodromies $\gamma_{L}$, $\gamma_{R}$
of $g_{L_1}$ and $g_{R_1}$:
\bea
\label{lrgam}
g_{L_1}(y+2\pi)=g_{L_1}(y)\gamma_L\\ \nonumber
g_{R_1}(y+2\pi)=g_{R_1}(y)\gamma_R
\eea
and
\be\label{mondelr}
 \gamma_R^{-1}\gamma_L=m_ah_a=d_a
\ee
It is instructive to compare (\ref{mondelr}) to (\ref{monl}) and (\ref{monr}).
We have seen in section 1, that in the absence of defect left and right monodromies
are equal, whereas presence of defect creates relative shift between them equal to the 
defect conjugacy class.
The symplectic form now is:
\be\label{sympdef}
\Omega^{\rm def1}={k\over 4\pi}\left[\int_0^a\Pi(g_1)dx+\int_a^{2\pi}\Pi(g_2)dx-\varpi(g_1(0,a),g_2(0,a))\right]
\ee
The conditions (\ref{dpi}) and (\ref{extdef}) imply that
\be
\delta\Omega^{\rm def1}=0
\ee
Substituting  (\ref{denlr}) in (\ref{sympdef}) one obtains:
\bea
{4\pi \over k}\Omega^{\rm def1}=\int_0^a{\rm tr}(g_{L_1}^{-1}\delta g_{L_1}\partial_x(g_{L_1}^{-1}\delta g_{L_1}))dx
-\int_0^a{\rm tr}(g_{R_1}^{-1}\delta g_{R_1}\partial_x(g_{R_1}^{-1}\delta g_{R_1}))dx\\ \nonumber
+\int_a^{2\pi}{\rm tr}(g_{L_2}^{-1}\delta g_{L_2}\partial_x(g_{L_2}^{-1}\delta g_{L_2}))dx
-\int_a^{2\pi}{\rm tr}(g_{R_2}^{-1}\delta g_{R_2}\partial_x(g_{R_2}^{-1}\delta g_{R_2}))dx\\ \nonumber
+{\rm tr}(g_{L_1}^{-1}\delta g_{L_1}g_{R_1}^{-1}\delta g_{R_1})(a)-
{\rm tr}(g_{L_1}^{-1}\delta g_{L_1}g_{R_1}^{-1}\delta g_{R_1})(0)\\ \nonumber
+{\rm tr}(g_{L_2}^{-1}\delta g_{L_2}g_{R_2}^{-1}\delta g_{R_2})(2\pi)-
{\rm tr}(g_{L_2}^{-1}\delta g_{L_2}g_{R_2}^{-1}\delta g_{R_2})(a)-\varpi(g_1(0,a),g_2(0,a))
\eea

Using (\ref{lra}) and (\ref{usfor}) one can check that
\bea
\label{varpi}
&&-\varpi(g_1(0,a),g_2(0,a))-{\rm tr}(g_{L_2}^{-1}\delta g_{L_2}g_{R_2}^{-1}\delta g_{R_2})(a)
+{\rm tr}(g_{L_1}^{-1}\delta g_{L_1}g_{R_1}^{-1}\delta g_{R_1})(a)\\ \nonumber
&&+{\rm tr}(h_a^{-1}\delta h_ag_{L_1}\delta g_{L_1})(a)+{\rm tr}(\delta m_am_a^{-1}g_{R_1}^{-1}\delta g_{R_1})(a)
=-\omega_{\mu_a}(d_a)+{\rm tr}(\delta h_ah_a^{-1}m_a^{-1}\delta m_a)
\eea
and
\be\label{gl2}
{\rm tr}(g_{L_2}^{-1}\delta g_{L_2}\partial_x(g_{L_2}^{-1}\delta g_{L_2}))
={\rm tr}(g_{L_1}^{-1}\delta g_{L_1}\partial_x(g_{L_1}^{-1}\delta g_{L_1}))
-\partial_x({\rm tr}(h_a^{-1}\delta h_ag_{L_1}\delta g_{L_1}))
\ee
\be\label{gr2}
{\rm tr}(g_{R_2}^{-1}\delta g_{R_2}\partial_x(g_{R_2}^{-1}\delta g_{R_2}))=
{\rm tr}(g_{R_1}^{-1}\delta g_{R_1}\partial_x(g_{R_1}^{-1}\delta g_{R_1}))+
\partial_x({\rm tr}(\delta m_am_a^{-1}g_{R_1}^{-1}\delta g_{R_1}))
\ee
Collecting all we get:
\bea
&&{4\pi \over k}\Omega^{\rm def1}=\\ \nonumber
&&\int_0^{2\pi}{\rm tr}(g_{L_1}^{-1}\delta g_{L_1}\partial_x(g_{L_1}^{-1}\delta g_{L_1}))dx
-\int_0^{2\pi}{\rm tr}(g_{R_1}^{-1}\delta g_{R_1}\partial_x(g_{R_1}^{-1}\delta g_{R_1}))dx\\ \nonumber
&&+{\rm tr}(\delta h_ah_a^{-1}m_a^{-1}\delta m_a)-{\rm tr}(h_a^{-1}\delta h_ag_{L_1}\delta g_{L_1})(2\pi)
-{\rm tr}(\delta m_am_a^{-1}g_{R_1}^{-1}\delta g_{R_1})(2\pi)\\ \nonumber
&&-{\rm tr}(g_{L_1}^{-1}\delta g_{L_1}g_{R_1}^{-1}\delta g_{R_1})(0)+{\rm tr}(g_{L_2}^{-1}\delta g_{L_2}g_{R_2}^{-1}\delta g_{R_2})(2\pi)-\omega_{\mu_a}(d_a)
\eea
Note that dependence on the insertion point $a$ is completely dropped. This reflects  topological nature of the defect.
Using (\ref{lrgam}) and (\ref{mondelr}) we derive 
\bea
\Omega^{\rm def1}=\Omega_L-\Omega_R+{k\over 4\pi}{\rm tr}(\delta\gamma_R\gamma_R^{-1}
\delta\gamma_L\gamma_L^{-1})-{k\over 4\pi}\omega_{\mu_a}(d_a)
\eea
Finally using decompositions of $g_{L_1}$ and $g_{R_1}$ (\ref{decomp}):
\be
g_{L_1}=h_Le^{i\tau_Lx/k}g_0 \;\;\; {\rm and} \;\;\; g_{R_1}=h_Re^{i\tau_Rx/k}f_0
\ee
with
$\gamma_L$ and $\gamma_R$:
\be\label{glrd}
\gamma_L=g_0e^{2i\pi\tau_L/k}g_0^{-1}\;\;\; {\rm and} \;\;\;\gamma_R=f_0e^{2i\pi\tau_R/k}f_0^{-1}
\ee
and the corresponding decomposition of $\Omega_{L,R}$ (\ref{omdecom}) we arrive at
\be\label{deff1}
\Omega^{\rm def1}=\Omega^{LG}(h_L,\tau_L)-\Omega^{LG}(h_R,\tau_R)+\Omega^{\rm defline}
\ee
\bea\label{omdefline1}
\Omega^{\rm defline1}=
{\rm tr}[(i\delta\tau_L)g_0^{-1}\delta g_0]-{\rm tr}[(i\delta\tau_R)f_0^{-1}\delta f_0]
\\ \nonumber
+{k\over 4\pi}\left[\omega_{\tau_L}(\gamma_L)-\omega_{\tau_R}(\gamma_R)
+{\rm tr}(\delta\gamma_R\gamma_R^{-1}
\delta\gamma_L\gamma_L^{-1})-\omega_{\mu_a}(d_a)\right]
\eea
Remembering (\ref{mondelr}) we see that defect phase space is
\be
{\cal P}^{\rm def1}=(d_a, g_0, \tau_L ,f_0,\tau_R)/G
\ee
subject to relation $\gamma_R^{-1}\gamma_L=d_a$ with $\gamma_L$ and $\gamma_R$
given by (\ref{glrd}). 
This is moduli space of flat connections on sphere with three Wilson lines ${\cal P}_{S^2_{3,0}}$.
The form (\ref{omdefline1}) coincides with (\ref{om3}) with terms (\ref{modterm}).
Comparing  (\ref{deff1}) with (\ref{decsn}) for $n=1$ and $m=2$,
we see that symplectic phase space of the WZW model on circle with one defect
coincides with that of CS on annulus with one Wilson line.

Let us briefly present the case of the two defects insertion.
 
Now let us put two defect lines, one at point $x=a$, and the second at point $x=b$,
again both in parallel to time line.
In this situation the world-sheet separated on three region, $\Sigma_1$,
$\Sigma_2$ and $\Sigma_3$, and correspondingly the WZW model is defined by three maps
$g_1$, $g_2$ and $g_3$.
At each point should be satisfied defect conditions (\ref{defcur}), bringing
as before to the following solution of the equations of motion:
\bea
g_{L_2}(y)=g_{L_1}(y)h_a^{-1},\hspace{2cm} g_{L_3}(y)=g_{L_2}(y)h_b^{-1}\\
g_{R_2}(y)=g_{R_1}(y)m_a, \hspace{2cm} g_{R_3}(y)=g_{R_2}(y)m_b
\eea
\be
m_ah_a=d_a\in {\cal C}_{\mu_a}, \hspace{2cm} m_bh_b=d_b\in {\cal C}_{\mu_b}
\ee
Requiring the condition of closedness of string 
\be
g_3(2\pi)=g_1(0)
\ee
brings to the following constraint on monodromies:
\bea
g_{L_1}(y+2\pi)=g_{L_1}(y)\gamma_L\\
g_{R_1}(y+2\pi)=g_{R_1}(y)\gamma_R
\eea
\be
\gamma_R^{-1}\gamma_L=m_am_bh_bh_a=m_ad_bm_a^{-1}d_a=\tilde{d_b}d_a
\ee
Note that relative shift between monodromies is equal to product of defect conjugacy classes. 
The symplectic form is:
\be
\Omega^{\rm def2}={k\over 4\pi}\left[\int_0^a\Pi(g_1)dx+\int_a^b\Pi(g_2)dx+\int_b^{2\pi}\Pi(g_3)dx-\varpi(g_1(0,a),g_2(0,a))-\varpi(g_2(0,b),g_3(0,b))\right]
\ee

Repeating the same steps as before we obtain:
\be\label{deff2}
\Omega^{\rm def2}=\Omega^{LG}(h_L,\tau_L)-\Omega^{LG}(h_R,\tau_R)+\Omega^{\rm defline2}
\ee
\bea\label{omdefline2}
&&\Omega^{\rm defline2}=
{\rm tr}[(i\delta\tau_L)g_0^{-1}\delta g_0]-{\rm tr}[(i\delta\tau_R)f_0^{-1}\delta f_0]
\\ \nonumber
&&+{k\over 4\pi}\left[\omega_{\tau_L}(\gamma_L)-\omega_{\tau_R}(\gamma_R)
-\omega_{\mu_a}(d_a)-\omega_{\mu_b}(\tilde{d}_b)
+{\rm tr}(\delta\gamma_R\gamma_R^{-1}\delta\gamma_L\gamma_L^{-1})+
{\rm tr}(\tilde{d}_b^{-1}\delta \tilde{d}_b\delta d_a d_a^{-1})\right]
\eea
The defect phase space now is
\be
{\cal P}^{\rm def2}=(d_a,\tilde{d}_b, g_0, \tau_L ,f_0,\tau_R)/G
\ee
subject to relation $\gamma_R^{-1}\gamma_L=\tilde{d_b}d_a$, 
with $\gamma_L$ and $\gamma_R$
given by (\ref{glrd}). This is phase space (\ref{moduli}) for $n=4$.
The form (\ref{omdefline2}) coincides with (\ref{om4a}) with terms (\ref{modterm}).
Comparing  (\ref{deff2}) with (\ref{decsn}) for $n=2$ and $m=2$,
we see that symplectic phase space of the WZW model on circle with two defects
coincides with that of CS on annulus with two Wilson lines.

These two examples can be easily generalized to the insertion of $N$ defects.

From these examples one can conclude that the defect phase space of the WZW 
model with $N$ defects insertion is
\be\label{defN}
{\cal P}^{\rm defN}=(d_1,\ldots d_N, g_0, \tau_L ,f_0,\tau_R)/G
\ee
subject to relation $\gamma_R^{-1}\gamma_L=\prod_{i=1}^Nd_i$, $d_i\in {\cal C}_{\mu_i}$ 
where ${\cal C}_{\mu_i}=\beta_ie^{2i\pi\mu_i/k}\beta_i^{-1}$,
with $\gamma_L$ and $\gamma_R$
given by (\ref{glrd}). This is phase space (\ref{moduli}) for $n=N+2$.
We see that defect fusion rule (\ref{dds}) corresponds in the classical picture
to the multiplication of the corresponding conjugacy classes.
By the cumbersome but straightforward calculation we can again check that the symplectic form on the defect phase space 
(\ref{defN}) is equal to symplectic form on the moduli space of flat connections on sphere
with $N+2$ sources
$\Omega_{S^2_{N+2,0}}$.

At the moment it is clear to author how to derive this result
 case by case by brute force calculation. More general understanding is desirable.

\subsection{Defects in open string}
In this section we consider WZW model with defect on strip.
Assume again that we have defect at point $x=a$ in parallel to the time line.
The strip is divided to two parts with fields $g_1$ and $g_2$.
We should impose here boundary conditions at $x=0$ on $g_1$, 
requiring 
\be\label{g1t}
g_1(t,0)\in {\cal C}_{\mu_0}=\beta_0e^{2i\pi\mu_0/k}\beta^{-1}_0,\;\;\; \beta_0\in G
\ee
then defect condition
at $x=a$, requiring 
\be\label{g1g2t}
g_1g_2^{-1}(t,a)\in {\cal C}_{\mu_a}=\beta_ae^{2i\pi\mu_a/k}\beta^{-1}_a,\;\;\; \beta_a\in G
\ee
and finally boundary condition at $x=\pi$ on $g_2$,
requiring
\be\label{g2tp}
g_2(t,\pi)\in {\cal C}_{\mu_{\pi}}=\beta_{\pi}e^{2i\pi\mu_{\pi}/k}\beta^{-1}_{\pi},\;\;\; \beta_{\pi}\in G
\ee
Equations (\ref{g1t}) and (\ref{g1g2t}) as before yield:
\be\label{r1l1}
g_{R_1}(y)=g_{L_1}(-y)h_0^{-1}
\ee
\be
g_1(0,t)=g_{L_1}(t)g_{R_1}^{-1}(-t)=g_{L_1}(t)h_0g_{L_1}^{-1}(t)
\ee
\be
h_0\in {\cal C}_{\mu_0}
\ee

\bea\label{l2l1}
g_{L_2}(y)=g_{L_1}(y)h_a^{-1}\\ \nonumber
g_{R_2}(y)=g_{R_1}(y)m_a
\eea
\be
m_ah_a=d_a\in {\cal C}_{\mu_a}
\ee
To solve the last boundary condition (\ref{g2tp}) we assume that $g_{L_1}$ has monodromy matrix $\gamma$:
\be\label{gam1}
g_{L_1}(y+2\pi)=g_{L_1}(y)\gamma
\ee
Using (\ref{r1l1}) and (\ref{l2l1}) one obtains:
\be\label{gl2g}
g_{L_2}(y+2\pi)=g_{L_2}(y)h_a\gamma h_a^{-1}
\ee
\be\label{gr2g}
g_{R_2}(y)=g_{L_2}(-y)h_ah_0^{-1}m_a
\ee
Equations (\ref{gl2g}) and (\ref{gr2g}) imply
\be
g_2(\pi,t)=g_{L_2}(\pi+t)g_{R_2}^{-1}(\pi-t)=g_{L_2}(-\pi+t)h_a\gamma h_a^{-1}m_a^{-1}h_0h_a^{-1}g_{L_2}^{-1}(-\pi+t)
\ee
To satisfy (\ref{g2tp}) one should require
\be\label{bd}
\gamma h_a^{-1}m_a^{-1}h_0=\gamma d_a^{-1}h_0=h_{\pi}\in {\cal C}_{\mu_{\pi}}
\ee
It is again instructive to compare (\ref{bd}) to (\ref{b}).
We see that presence of defect again requires to include defect conjugacy class.
This is classical analogue of the defect-boundary fusion (\ref{dbs}).
The symplectic form is
\be
\Omega^{\rm strip-def}={k\over 4\pi}\left[\int_0^a\Pi(g_1)dx+\int_a^{\pi}\Pi(g_2)dx-\varpi(g_1(0,a),g_2(0,a))+\omega_{\mu_0}(g_1(0,0))
-\omega_{\mu_{\pi}}(g_2(0,\pi))\right]
\ee

Executing the same steps as in previous sections we finally obtain:
\be\label{strdef}
\Omega^{\rm strip-def}=\Omega^{LG}(h,\tau)+\Omega^{\rm bndry-def}
\ee
where
\bea\label{bndrdef}
\Omega^{\rm bndry-def}={\rm tr}[(i\delta\tau)g_0^{-1}\delta g_0]+
{k\over 4\pi}\left[
\omega_{\tau}(\gamma)+
\omega_{\mu_0}(h_0)-\omega_{\mu_{\pi}}(h_{\pi})-\omega_{\mu_a}(d_a)\right.\\ \nonumber
+{\rm tr}(d_a^{-1}\delta h_0h_0^{-1}d_a\gamma^{-1}\delta\gamma)
+{\rm tr}(\gamma^{-1}\delta\gamma d_a^{-1}\delta d_a)
+\left.{\rm tr}(\delta d_a d_a^{-1}\delta h_0h_0^{-1})\right]
\eea
The boundary-defect phase space is
\be
{\cal P}^{\rm bndry-def}=(h_0,h_{\pi},d_a, g_0,\tau)/G
\ee
subject to relation $\gamma d_a^{-1}h_0=h_{\pi}$, where $h_0\in {\cal C}_{\mu_0}$,
$h_{\pi}\in {\cal C}_{\mu_\pi}$, $d_a\in {\cal C}_{\mu_a}$, and $\gamma=g_0e^{2i\pi\tau/k}g_0^{-1}$.
This is phase space (\ref{moduli}) for $n=4$.
We see that (\ref{bndrdef}) coincides with (\ref{om4b}).
Comparing  (\ref{strdef}) with (\ref{decsn}) for $n=3$ and $m=1$,
we see that symplectic phase space of the WZW model on strip with one defect
coincides with that of CS theory on disc with three Wilson lines.
As we explained in the previous section, consideration here can be generalized to the case of insertion
of the arbitrary number of the defect lines as well, yielding the symplectomorphism between
phase space of the WZW model on strip with $N$ defects 
with that of CS theory on disc with $N+2$ Wilson lines.

\section{Permutation branes}
Maximally symmetric permutation branes on two-fold product 
of the WZW models $G\times G$ is defined as boundary conditions satisfying the relations:
\be\label{curboundp}
J_{L_1}=-J_{R_2}|_{\partial M}
\ee
and
\be\label{curboundpp}
J_{L_2}=-J_{R_1}|_{\partial M}
\ee
Here label $1$ and $2$ refer two the first and the second copy.
It was shown in \cite{FigueroaO'Farrill:2000ei} that conditions 
(\ref{curboundp}) and (\ref{curboundpp}) imply that values of $g_1$ and $g_2$ 
on the boundary constrained by the relation:
\be\label{perg}
g_1g_2|_{\partial M}=\tilde{F}\in {\cal C}_{\mu}=\beta e^{2i\pi\mu/k}\beta^{-1},\;\;\;\; \beta\in G
\ee
It is shown in \cite{Sarkissian:2003yw} that in the Lagrangian approach  to the boundary WZW model
as explained in the section 4, the permutation branes correspond to Lagrangian:
\be\label{actdefp}
S=S^{\rm bulk}(g_1)+S^{\rm bulk}(g_2)-{k\over 4\pi}\int_D\omega_{\cal P}(g_1,g_2)
\ee
where
\be\label{varmarp}
\omega_{\cal P}(g_1,g_2)=\omega_{\mu}(\tilde{F})+{\rm Tr}(g_1^{-1}dg_1dg_2g_2^{-1})
\ee
The form (\ref{varmarp}) satisfies the equation:
\be\label{extdefp}
d\omega_{\cal P}(g_1,g_2)=\omega^{WZ}(g_1)|_{\rm boundary}+\omega^{WZ}(g_2)|_{\rm boundary}
\ee
Equation (\ref{extdefp}) guarantees that the action (\ref{actdefp}) is well defined.
Consider now two-fold product on a strip with boundary conditions (\ref{curboundp}) and (\ref{curboundpp})
imposed at points $x=0$ and $x=\pi$.
It is possible to show that equations of motions (\ref{eqmot}) with these boundary conditions can be solved 
by (\ref{bulkdec}) on bulk for both fields with $g_{L_1}$,$g_{R_1}$,$g_{L_2}$,$g_{R_2}$
satisfying:
\be\label{g1}
g_{L_1}(y+2\pi)=g_{L_1}(y)\gamma_1
\ee
\be\label{g2}
g_{L_2}(y+2\pi)=g_{L_2}(y)\gamma_2
\ee
\be\label{h0}
g_{R_2}(y)=g_{L_1}(-y)h_0^{-1}
\ee
\be\label{m0}
g_{R_1}(y)=g_{L_2}(-y)m_0^{-1}
\ee

From (\ref{h0}), (\ref{m0}) we obtain:

\be
\tilde{F}(0,t)=g_{L_1}(t)m_0h_0g_{L_1}^{-1}(t)
\ee

Therefore to be in agreement with (\ref{perg}) we should require:
\be
m_0h_0=p_0\in {\cal C}_{\mu_0}
\ee

Equations (\ref{g1}) and (\ref{g2}) further imply 

\be
\tilde{F}(\pi,t)=g_{L_1}(-\pi+t)\gamma_1m_0\gamma_2h_0g_{L_1}^{-1}(-\pi+t)
\ee
Therefore we additionally should require:
\be\label{gcon}
\gamma_1m_0\gamma_2h_0=\gamma_1p_0h_0^{-1}\gamma_2h_0=\gamma_1p_0\tilde{\gamma}_2=p_{\pi}\in {\cal C}_{\mu_{\pi}}
\ee
where
\be
\tilde{\gamma}_2=h_0^{-1}\gamma_2h_0
\ee

The symplectic form corresponding to the action (\ref{actdefp}) on the strip is
\be
\Omega_{\cal P}={k\over 4\pi}\left[\int_0^{\pi}(\Pi(g_1)+\Pi(g_2))dx+\omega_{\cal P}(g_1(0,0),g_2(0,0))-
\omega_{\cal P}(g_1(0,\pi),g_2(0,\pi))\right]
\ee

Repeating the same steps as explained in the previous sections we obtain:

\be\label{omperm}
\Omega_{\cal P}=\Omega^{LG}(h_1,\tau_1,)+\Omega^{LG}(h_2,\tau_2,)+\Omega^{\rm bndry-perm}
\ee
\bea\label{bndryperm}
\Omega^{\rm bndry-perm}={\rm tr}[(i\delta\tau_1)g_0^{-1}\delta g_0]
+{\rm tr}[(i\delta\tau_2)f_0^{-1}\delta f_0]\\ \nonumber
+{k\over 4\pi}\left[\omega_{\tau_1}(\gamma_1)+\omega_{\tau_2}(\tilde{\gamma}_2)+\omega_{\mu_0}(p_0)-
\omega_{\mu_{\pi}}(p_{\pi})\right.\\ \nonumber
-\left.{\rm tr}(p_0^{-1}\delta p_0\delta \tilde{\gamma}_2\tilde{\gamma}_2^{-1})
-{\rm tr}(\gamma_1^{-1}\delta\gamma_1\delta p_0p_0^{-1})-{\rm tr}(p_0^{-1}\gamma_1^{-1}\delta\gamma_1p_0\delta \tilde{\gamma}_2\tilde{\gamma}_2^{-1})\right]
\eea
Comparing (\ref{gcon}) to (\ref{moduli}), (\ref{bndryperm}) to (\ref{om4b}), and finally
(\ref{omperm}) with (\ref{decsn}) for $n=2$ and $m=2$,
we see that symplectic phase space of the WZW model $G\times G$ on strip with boundary conditions
specified by permutation branes 
coincides with that of CS on annulus with two Wilson lines. 
The generalization to the case of permutation branes on $N$-fold product is again
cumbersome but straightforward.

\newpage

\noindent {\bf Acknowledgements} \\[1pt]
I am grateful to Christoph Schweigert for useful discussions
and comments on the paper. \\
Author received partial support from the Collaborative Research Centre 
676 ``Particles, Strings and the Early Universe - the Structure of Matter and 
Space-Time''.

\end{document}